\documentclass[nofootinbib,amsmath,amssymb,10pt,eqsecnum, twocolumn]{revtex4-1}
\usepackage{graphicx, amsmath, amsmath, amssymb, bm}
\usepackage{tabularx}
\usepackage[caption=false]{subfig}
\usepackage[breaklinks, colorlinks, citecolor=blue]{hyperref}
\usepackage{color}
\usepackage{float}
\usepackage{natbib,hyperref,ifthen}

\newcommand{\beq}{\begin{equation}}
\newcommand{\eeq}{\end{equation}}

\def\ee{\end{equation}}
\def\bea{\begin{eqnarray}}
\def\eea{\end{eqnarray}}
\def\bse{\begin{subequations}}

\newcolumntype{Y}{>{\centering\arraybackslash}X}
\newcolumntype{b}{>{\centering\arraybackslash}X}
\newcolumntype{s}{>{\centering\arraybackslash\hsize=.333\hsize}X}

\begin{document}

\title{Accelerated Bayesian inference using deep learning}

\author{Adam Moss} \email{adam.moss@nottingham.ac.uk}
\affiliation{School of Physics \& Astronomy\\
University of Nottingham,
Nottingham, NG7 2RD, England}

\date{\today}

\begin{abstract}
   We present a novel Bayesian inference tool that uses a neural network to parameterise efficient Markov Chain Monte-Carlo (MCMC) proposals.  The target distribution is first transformed into a diagonal, unit variance Gaussian by a series of non-linear, invertible, and non-volume preserving flows. Neural networks are extremely expressive, and can transform complex targets to  a simple latent representation. Efficient proposals can then be made in this space, and we demonstrate a high degree of mixing on several challenging distributions. Parameter space can naturally be split into a block diagonal speed hierarchy, allowing for fast exploration of subspaces where it is inexpensive to evaluate the likelihood.  Using this method, we develop a nested MCMC sampler to perform Bayesian inference  and model comparison, finding excellent performance on highly curved and multi-modal analytic likelihoods.  We also test it on {\em Planck} 2015 data, showing accurate parameter constraints, and calculate the evidence for simple one-parameter  extensions to LCDM in $\sim20$ dimensional parameter space. Our method has wide applicability to a range of problems in astronomy and cosmology.
\end{abstract}

\maketitle


\section{Introduction}

In the last few years, we have witnessed a revolution in machine learning. The use of deep neural networks (NNs) has become widespread due to increased computational power, the availability of large datasets, and their ability to solve problems previously deemed intractable (see~\cite{0483bd9444a348c8b59d54a190839ec9} for an overview).  Deep learning is particularly suited to the era of data driven astronomy and cosmology, but so far applications have mainly focused on supervised learning tasks such as classification and regression.

Bayesian inference is now a standard technique, with codes such as {\tt CosmoMC}~\cite{2002PhRvD..66j3511L},  {\tt CosmoSIS}~\cite{2015A&C....12...45Z},  {\tt Emcee}~\cite{2013ascl.soft03002F},  {\tt MontePython}~\cite{Audren:2012wb},  {\tt MultiNest}~\cite{2008MNRAS.384..449F, 2009MNRAS.398.1601F, 2013arXiv1306.2144F} and {\tt PolyChord}~\cite{2015MNRAS.453.4384H, 2017arXiv170403459H} used for parameter estimation and model selection. Many of these use Metropolis-Hastings Markov Chain Monte Carlo (MCMC) to draw samples from the posterior distribution. The dimensionality of the problem can be high, with $\gtrsim20$  parameters for the {\em Planck} likelihood (including nuisance parameters), and potentially  more   for future large-scale surveys. For some problems the likelihood can be non-Gaussian (e.g. with curving degeneracies) and/or multi-modal,  making conventional MCMC techniques inefficient and liable to miss regions of parameter space. Likelihoods can also be expensive to calculate, so minimizing the total number of evaluations is advantageous. This was the motivation behind {\tt BAMBI}~\cite{2012MNRAS.421..169G, pybambi}, which used a neural network to approximate the likelihood function where possible. Efficient exploration of the posterior distribution  is crucial for Bayesian inference and model selection. 

A good proposal function is vital to fully explore the distribution and ensure convergence of the chain. Deep learning has recently been shown to accelerate MCMC sampling by using NNs to parameterise efficient proposals that maintain detailed balance (see e.g.~\cite{2017arXiv171109268L, 2017arXiv170607561S}). The proposal function can be trained, for example, to minimise the autocorrelation length of the chain. Some of these methods (e.g. generalisations of Hamiltonian Monte Carlo) exploit the gradient of the target, and analytic gradients are often not available for astronomical or cosmological models. The training time can also be prohibitive, offsetting the gain in efficiency when sampling. In this work we use a NN to transform the likelihood  to a simpler representation, which requires no gradient information and is very fast to train. This approach is inspired by {\em representation learning}, which hypotheses that deep NNs  have the potential to yield representation spaces in which Markov chains mix faster~\cite{2012arXiv1206.5538B, 2012arXiv1207.4404B}.

The idea of optimising the proposal function originated in~\cite{Gelman96, haario2001adaptive}, which suggested using a normal distribution $\mathcal{N} ({\bf x},  2.38^2 \, {\bf \Sigma} / d)$, where ${\bf \Sigma}$ is the covariance matrix, ${\bf x}$ the current state  and $d$ the dimension. The factor of $2.38^2 /d$ was shown to minimise the autocorrelation length of the chain.  This is equivalent to transforming variables by ${\bf L}^{-1} {\bf x}$, where  ${\bf L}$ is a lower triangular matrix defined by the Cholesky decomposition of the covariance matrix, ${\bf \Sigma} = {\bf L} {\bf L}^T$ . Proposals can then be made using  the diagonal normal distribution  $\mathcal{N} ({\bf x},  2.38^2 \, {\bf I} / d)$, where ${\bf I}$ is the identity matrix. Standard practice in cosmological parameter estimation is to use an initial set of samples to estimate the covariance matrix, and make proposals using the Cholesky decomposition. Linear transformations are also used in affine invariant MCMC methods~\cite{2010CAMCS...5...65G, 2013ascl.soft03002F}.

This transformation works well when samples are well described by their covariance matrix,  but can become inefficient for more complex  distributions. In this paper we use a  NN to parameterise more expressive transformations that are suitable for curved and multi-modal  targets. The NN learns a non-linear, invertible, and non-volume preserving  mapping between data and  a  simpler latent space via maximum-likelihood,  and proposals are then made in this space. We apply our method to nested sampling, a commonly used tool to perform Bayesian inference and model comparison, although it could easily be incorporated into other MCMC frameworks. A major challenge in nested sampling is drawing new samples from a constrained target distribution, and we show that NNs can lead to improved performance over  existing rejection and MCMC based approaches.

The structure of the paper is as follows. In section~\ref{sec:nested} we provide a short overview of nested sampling. In section~\ref{sec:network}  we show how NNs can be trained to transform complex target data to a simpler latent space. We give further details of our algorithm in section~\ref{sec:method}, and demonstrate results on both analytic likelihoods and {\em Planck} data in section~\ref{sec:results}. We conclude and provide an outlook for future work in section~\ref{sec:conclusions}.


\section{ Nested Sampling } \label{sec:nested}

The nested sampling algorithm~\cite{2004AIPC..735..395S, skilling2006nested}  was developed to accurately calculate the Bayesian evidence (or marginal likelihood). From Bayes' theorem, the posterior distribution of a set of parameters ${\bf x}$, given data ${\bf d}$ and model $M$ is

\begin{equation}
p ({\bf x} | {\bf d}, M) = \frac{p ({\bf d}| {\bf x}, M) p({\bf x} | M)}{p({\bf d} | M)}\,,
\end{equation}
where $p ({\bf d}| {\bf x}, M) = \mathcal{L}({\bf x}) $ is the likelihood,  $p({\bf x} | M)= \pi({\bf x})$ is the prior and the normalising constant $p({\bf d} | M)$ is the evidence. This can be expressed as 
\begin{equation}  \label{eqn:evidence}
Z =  p({\bf d} | M) = \int  \mathcal{L}({\bf x})  \pi({\bf x}) d {\bf x}\,.
\end{equation}
Two competing models $M_1$ and $M_2$ can be compared by calculating the Bayes factor
\begin{equation}
B = \frac{p(M_1 | {\bf d})}{p(M_2 | {\bf d})} = \frac{\pi(M_1)}{\pi(M_2)}  \frac{p({\bf d} | M_1) }{p({\bf d} | M_2) }\,,
\end{equation}
which simplifies to the evidence ratio if the models have equal prior probability.

This integral~(\ref{eqn:evidence}) is typically hard to evaluate, but can be turned into a simpler 1-d integral by a change of variables~\cite{2004AIPC..735..395S}, 
\begin{equation} \label{eqn:nested}
X(\lambda) = \int_{\mathcal{L} ({\bf x}) > \lambda} \pi({\bf x}) d {\bf x}\,,  \quad Z = \int_0^1 \mathcal{L}(X)  d X\,,
\end{equation}
such that $X(\lambda)$ is the prior volume associated with a likelihood constant  $\mathcal{L} ({\bf x}) > \lambda$. Parameters are scaled to the unit hypercube, so that the prior is normalised, and $X$ takes values between 0 and 1.

The nested sampling algorithm first draws a set of $n_{\rm live}$ {\em live} points from $\pi({\bf x})$. The prior is typically  uniform or Gaussian, so is easy to sample from. At each iteration the point  with the lowest likelihood is replaced by a new sample drawn from the prior, with the condition that the new point has a higher likelihood. The discarded samples are termed {\em dead} points. For a sequence of decreasing $X$ values,
\begin{equation}
0 < X_M < \cdots < X_2 < X_1 < X_0 = 1\,,
\end{equation}
the integral in ($\ref{eqn:nested}$) can then be estimated using trapezoidal integration
\begin{equation}
Z = \sum_{i=1}^M \mathcal{L}_i w_i\,,
\end{equation}
where $\mathcal{L}_i = \mathcal{L} (X_i)$ and $w_i = \frac{1}{2} \left( X_{i-1} - X_{i+1} \right)$. The error in $\log Z$ can be estimated by $\sqrt{H/n_{\rm live}}$, where $H$ is the negative relative entropy~\cite{2008MNRAS.384..449F} 
\begin{equation} \label{eqn:zerr}
H =  \sum_{i=1}^M \frac{\mathcal{L}_i w_i }{Z} \log{\frac{\mathcal{L}_i}{Z}}\,.
\end{equation}

Nested sampling can  also perform posterior inference by using the sequence of dead points (and the current set of live points), and assigning a  weight $p_i$ to the $i$th point
\begin{equation}
p_i = \frac{\mathcal{L}_i w_i }{Z}\,.
\end{equation}

The main difficulty with nested sampling is drawing a new, independent sample subject to a hard likelihood constraint.  This can be achieved by rejection sampling, using an envelope function that encloses the current set of live points. For Gaussian likelihoods, it is most efficient to use a ellipsoidal envelope~\cite{2006ApJ...638L..51M}, and for multi-modal distributions a set of (possibly) overlapping ellipsoids can be drawn around clusters of live points~\cite{2008MNRAS.384..449F, 2009MNRAS.398.1601F}. The disadvantage of rejection sampling is that the envelope function requires scaling by an enlargement factor to ensure it contains the entire iso-likelihood contour. This can lead to poor scaling with dimension and introduces a choice of user specified  hyper-parameter.

An alternative to rejection sampling  is MCMC. Starting from an existing live point,  a new, independent sample is obtained after performing a `sufficient' number of steps. MCMC scales better with dimension, but is not guaranteed to explore or fully mix the distribution, and is generally not well suited  if the likelihood is highly curved and/or multi-modal.  Variants of MCMC developed to cope with challenging targets include Galilean dynamics~\cite{2013AIPC.1553..106F}, diffusive sampling~\cite{2009arXiv0912.2380B} and slice sampling~\cite{2015MNRAS.453.4384H, 2017arXiv170403459H}. Some of these also use clustering algorithms to identify and sample from multi-modal distributions. The key point is that they all try and choose better proposals -- in our case we will use a neural network to try and learn one.


\section{ Neural Network Sampling } \label{sec:network}

\subsection{Non-volume preserving flows}

Initially, we have a set of data $\{ {\bf x}^i \}_{i=1}^N$, sampled from a target distribution ${\bf x} \sim p_X ({\bf x})$, where ${\bf x}$ has dimension $n_{\rm dim}$.  Latent variables are drawn from a simpler prior distribution ${\bf z} \sim p_Z ({\bf z})$, with the same dimension. Given a bijection $f: X \rightarrow Z$, the change of variables formula gives the distribution on $X$
\begin{equation} \label{eqn:change}
p_X({\bf x}) = p_Z(f{(\bf x})) \left| \det \frac{\partial f({\bf x})}{{\partial \bf x}} \right| \,.
\end{equation}
The inverse $f^{-1}{(\bf z})$ provides a mapping from latent space to real space. The bijection can be parameterised by a neural network, with trainable parameters $\theta$. Neural networks, however, are not generally invertible, and the Jacobian determinant in~(\ref{eqn:change}) is not easily tractable.

Non-volume preserving (NVP) flows, introduced in~\cite{2016arXiv160508803D}, can transform simple latent distributions into rich target distributions. They are invertible and allow for tractable Jacobians by using a specific architecture for the neural network. They exploit the fact that the determinant of a triangular matrix is the product of its diagonal terms. The NVP transformation is~\cite{2016arXiv160508803D}
\begin{equation}
{\bf x}^{\prime} = {\bf m} \odot {\bf x} + (1 - {\bf m}) \odot \Big({\bf x} \odot \exp\big(s_{\theta_s}({\bf m} \odot {\bf x})\big) + t_{\theta_t}({\bf m} \odot {\bf x})\Big)\,,
\end{equation}
where ${\bf m}$ is a  binary mask vector consisting of alternating 1's and 0's, $s_{\theta_s}$ and $t_{\theta_t}$ are separate (s)cale and (t)ranslation NN's with trainable parameters $\theta_s$ and $\theta_t$, and $\odot$ is the element-wise product. The transformation for $n_{\rm dim}=2$ with ${\bf m}= (1, 0)$, for example, is
\begin{eqnarray}
x_1^{\prime} &=&  x_1  \\ \nonumber 
x_2^{\prime} &=& x_2 \exp( s_{\theta_s} (x_1)) +  t_{\theta_t} (x_1)\,,
\end{eqnarray}
with Jacobian determinant $\exp( s_{\theta_s} (x_1)) $, and the inverse is
\begin{eqnarray}
x_1 &=&  x_1^{\prime}  \\ \nonumber 
x_2 &=& \left( x_2^{\prime} - t_{\theta_t} (x_1) \right) \exp( - s_{\theta_s} (x_1)) \,,
\end{eqnarray}
with Jacobian determinant $\exp(- s_{\theta_s} (x_1)) $. Note that $x_1$ is unmodified in this transformation.

A series of transformations can be composed into a {\em flow} by permuting components of the inputs in successive transformations, such that those modified in one transformation are left unchanged in the next.  This can be achieved by setting the mask in the next transformation to $1 - {\bf m}$, so that successive masks resemble a checkerboard pattern. The Jacobian determinant is still tractable, and is simply the product of each individual transformation. A  flow transforms a target to latent space, and an inverse flow transforms latent space to the target. Each transformation step of the flow has separate $s$ and $t$ networks.

Given a set of samples, the $s$ and $t$ networks can be trained by minimising the loss function
\begin{eqnarray} \label{eqn:loss}
L &=& - \sum_i^N \log p_X({\bf x}^i)\,  \\ \nonumber
&=& - \sum_i^N \log p_Z(f{(\bf x}^i))  + \log  \left| {\rm det} \frac{\partial f({\bf x}^i)}{{\partial \bf x}^i} \right| \,.
\end{eqnarray}
Parameters are updated by back-propagating the loss using gradient descent, and the minimum loss is equivalent to  the maximum-likelihood estimate of ${\theta_s}$ and ${\theta_t}$. 

\begin{figure}
\centering
\includegraphics[width=88mm, angle=0]{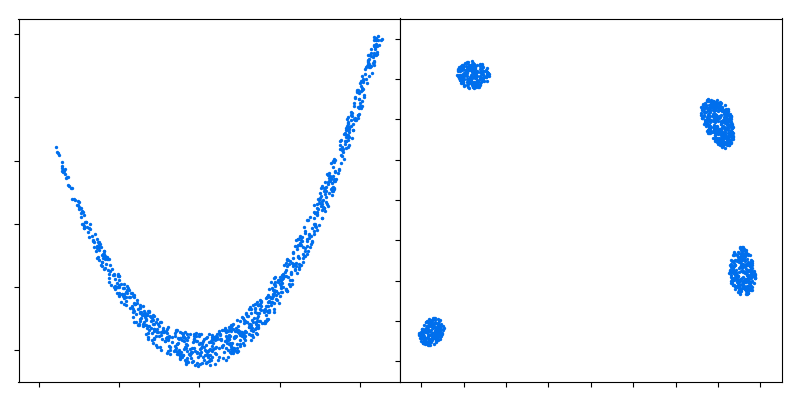}
\caption{\label{fig:flow_samples} Example data  ${\bf x} \sim p({\bf x})$ of 1000 samples used to train the network for (left) the $n_{\rm dim}=2$ Rosenbrock function, and (right) the Himmelblau function. }
\end{figure}

As an example, we fit flows to two datasets, shown in Fig~\ref{fig:flow_samples}. The initial data  ${\bf x} \sim p({\bf x})$ is obtained from the nested sampling of the 2 dimensional Rosenbrock and Himmelblau functions  (defined in section~\ref{sec:results}) for $n_{\rm live}=1000$ points, evaluated when the prior volume $X=0.02$. The target distribution is therefore uniform when $\mathcal{L} ({\bf x}) > \lambda_{\star}$, with $\lambda_{\star}$ defined by $X(\lambda_{\star})=0.02$, otherwise the probability density is zero. These functions are chosen as they are challenging examples of curved and multi-modal distributions.

We choose the prior distribution $p_Z$ to be a diagonal Gaussian with unit variance, i.e. ${\bf z} \sim \mathcal{N}(0, {\bf I})$. We use 4 successive transformations in the flow, each parameterised by a fully connected  $s$ and $t$ network with an input layer of dimension 2, two hidden layers of dimension 128, and an output layer of dimension 2.  We use rectified non-linear activation (ReLU) functions after the input and hidden layers in each network. 

\begin{figure*}
\centering
\includegraphics[width=\textwidth, angle=0]{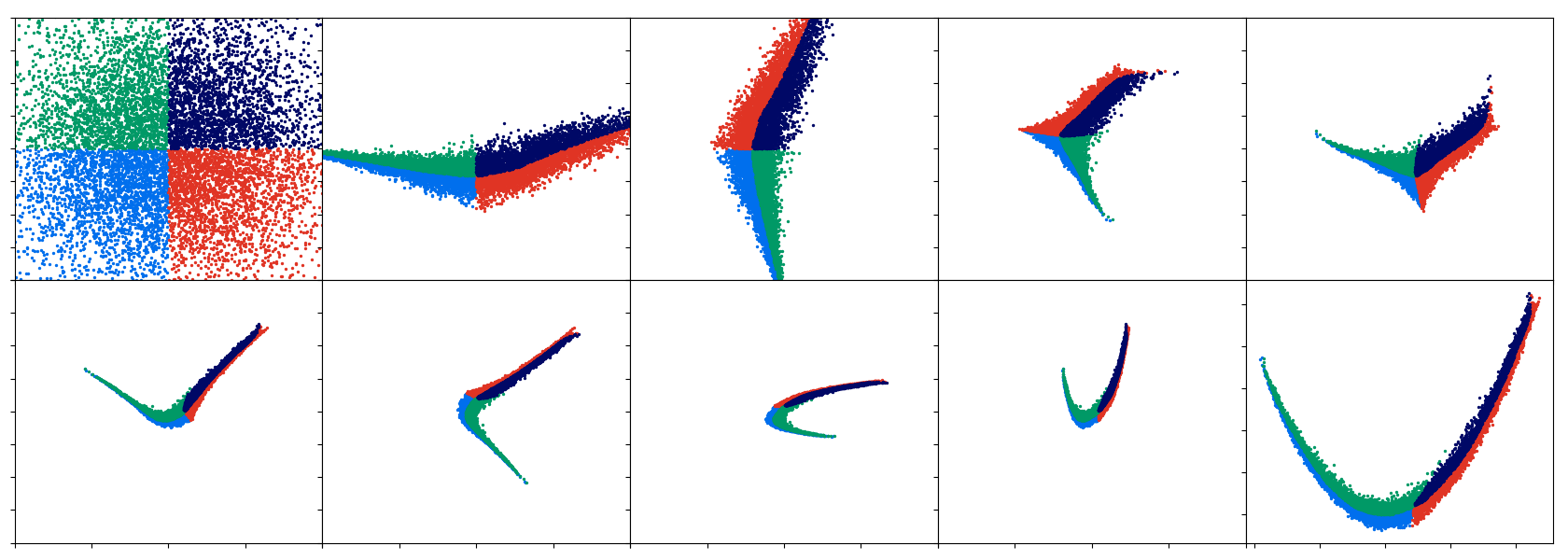}
\includegraphics[width=\textwidth, angle=0]{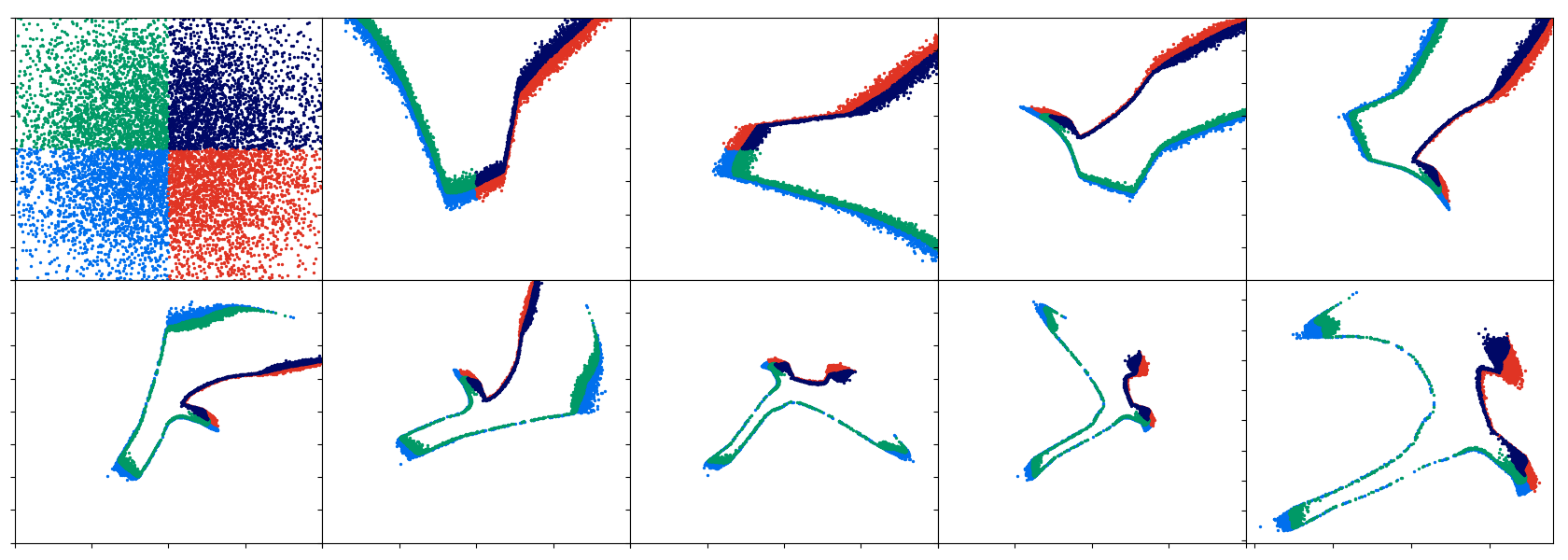}
\caption{\label{fig:flow} Inverse flow for (top) the  $n_{\rm dim}=2$ Rosenbrock function and (bottom) the Himmelblau function, trained on the initial data shown in Fig.~\ref{fig:flow_samples}. The inverse flow starts from latent space, ${\bf z} \sim \mathcal{N}(0, {\bf I})$, and each step consists of a transformation followed by a  permutation. The final image in each case ends at the target distribution and is zoomed in for clarity.}
\end{figure*}

The resulting inverse flows are shown in Fig.~\ref{fig:flow}, after training for 50 epochs (each epoch is a complete pass over training data). Each transformation step begins with an ($x_1$ dependent) scaling and translation of $x_2$  (the vertical axis in the plot), with $x_1$ unchanged. These are then permuted, and scaling and transformation operations are applied to (the now) $x_1$. The initial Gaussian can be flowed into the Rosenbrock function continuously, but narrow connecting ridges appear for the Himmelblau function. This is because it cannot continuously be deformed  into the target distribution. Nethertheless, the volume of these ridges is quite small compared to the region where the target probability density is non-zero.
 
If the network could fit the target distribution perfectly, it would be trivial to generate new, independent samples. One could simply sample from latent space  ${\bf z} \sim \mathcal{N}(0, {\bf I})$ and perform the inverse flow to obtain ${\bf x}$. In reality the fit isn't perfect, but we can use the learnt mapping to improve the efficiency of MCMC by making proposals in the simpler  latent space rather than data space.


\subsection{MCMC Sampling}

The acceptance probability required to maintain detailed balance  in a Metropolis-Hastings update is
\begin{equation} 
\alpha = {\rm min} \left(1, \frac{p_X({\bf x}^{\prime}) q ({\bf x} | {\bf x}^{\prime})}{p_X({\bf x}) q({\bf x}^{\prime}| {\bf x})  } \right)\,,
\end{equation}
where the proposal function $q({\bf x}^{\prime}| {\bf x}) $ is the conditional probability of state ${\bf x}^{\prime}$ given ${\bf x}$. If the proposal function is symmetric (e.g. a Gaussian with the same covariance matrix for each state) then  $q({\bf x}^{\prime}| {\bf x}) =  q({\bf x}| {\bf x}^{\prime}) $.  

For proposals made in latent space ${\bf z}$, the  acceptance probability must be modified by the Jacobian determinant to satisfy detailed balance
\begin{equation} 
\alpha = {\rm min} \left(1, \frac{p_X(f^{-1}({\bf z}^{\prime})) q ({\bf z} | {\bf z}^{\prime})  \left| \det \frac{\partial f^{-1}({\bf z}^{\prime})}{{\partial \bf z}^{\prime}} \right| }{p_X(f^{-1}({\bf z})) q({\bf z}^{\prime}| {\bf z})   \left| \det \frac{\partial f^{-1}({\bf z}^{})}{{\partial \bf z}^{}} \right| } \right)\,.
\end{equation}
Given the prior distribution of latent space is a diagonal, unit variance Gaussian, we use a symmetric proposal function
\begin{equation}
q ({\bf z}^{\prime} | {\bf z}^{}) = \mathcal{N} ({\bf z},  \sigma^2 \, {\bf I})\,,
\end{equation}
where $\sigma$ is a scaling parameter. Based on estimates of optimal proposals for Gaussian distributions~\cite{Gelman96, haario2001adaptive}, we tune $\sigma$ to give an acceptance rate of $50\%$ using the method in~\cite{2008MNRAS.384..449F},
\begin{equation} \label{eqn:proposalwidth}
\sigma \rightarrow \left\{ \begin{array}{ll}
         \sigma e^{1/N_{\rm a}} & \mbox{if $N_{\rm a} > N_{\rm r}$}\\
         \sigma e^{-1/N_{\rm r}} & \mbox{if $N_{\rm a} \le N_{\rm r}$}
\end{array} \right.,
\end{equation}
where $N_{\rm a}$ and $N_{\rm r}$  are the number of accepted and rejected samples in the current MCMC chain.

\begin{figure}
\centering
\includegraphics[width=88mm, angle=0]{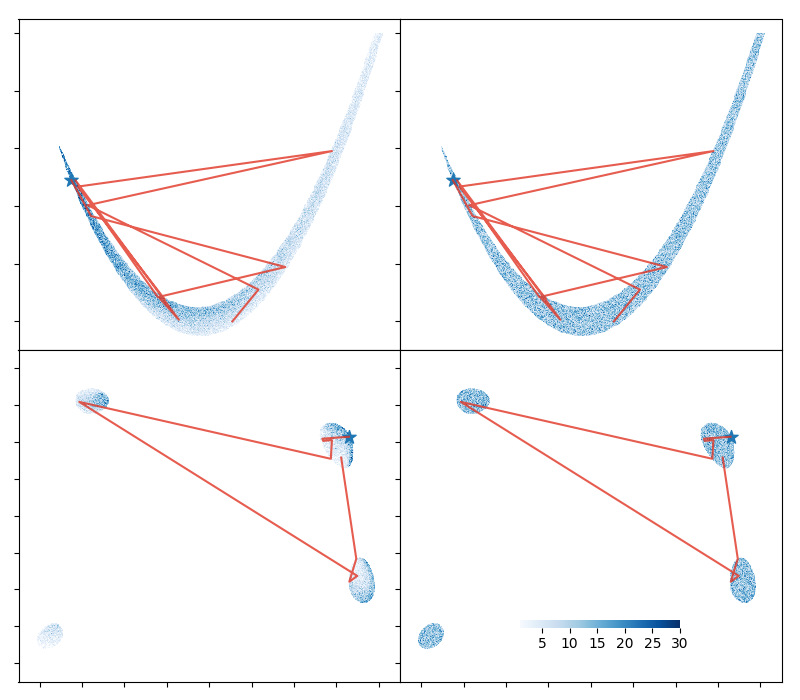}
\caption{\label{fig:flow_chains} MCMC chains for (top) the  $n_{\rm dim}=2$ Rosenbrock function and (bottom) the Himmelblau function.  On the left we show the histogram of samples after 5 MCMC iterations, and on the right after 20 MCMC iterations.  The initial position of all chains is indicated by a star. After 20 iterations the samples appear to have lost  all memory of where they began. We also show the first 20 proposed moves of an example chain.
 }
\end{figure}

To demonstrate this gives the correct distribution on $p_X$, we generate new samples for the flows fitted to the 2 dimensional Rosenbrock and Himmelblau functions. Starting from an existing sample ${\bf x}_{\rm init}$, we run a chain of length $N_{\rm c} = 1000$ and repeat this for  20,000 chains, each time choosing the {\em same} initial ${\bf x}_{\rm init}$. In Fig.~\ref{fig:flow_chains} we plot a histogram of the resulting samples after 5 and 20 MCMC iterations, also showing the first 20 proposed moves of an example chain (note that not all of these proposals are accepted). After only 5 iterations, the resulting distribution is non-uniform, with a higher probability near the initial point, but after 20 iterations the samples appear to have lost  all memory of where they began.  By sampling in latent space, the chain is able to take large steps in data space, even jumping directly between modes, with an overall acceptance rate in each case of around $40\%$. 

To quantify how many iterations are required to generate a new, independent sample, we calculate the effective sample size (ESS). Given a chain of $N_{\rm c}$ correlated samples $\{ {\bf x}^i \}_{i=1}^{N_{\rm c}}$, the ESS is
\begin{equation}
{\rm ESS} = \frac{N_{\rm c}}{1 + 2 \sum_{s=1}^{N_{\rm c}-1}(1 - s/N_{\rm c}) {\bm \rho}_s}\,,
\end{equation}
where  ${\bm \rho}_s$ is the autocorrelation of ${\bf x}$ at lag $s$. Since  there is an autocorrelation and ESS for each parameter,  we use the (worst-case) minimum ESS to set the chain length requirement. We use the following estimate for ${\bm \rho}_s$,
\begin{equation}
 \hat{\bm \rho}_s = \frac{1}{\hat{\bm \sigma}^2 (N_{\rm c} - s)} \sum_{n = s+1}^{N_{\rm c}} ({\bf x}^n - \hat{\bm \mu}) ({\bf x}^{n-s} - \hat{\bm \mu})\,,
\end{equation}
where $\hat{\bm \mu}$ and $\hat{\bm \sigma}^2$ are the mean and variance of the initial data. We truncate the sum over ${\bm \rho}_s$ when $ \hat{\bm \rho}_s  < 0.05$, as the estimate can become dominated by noise for large lags~\cite{carlin2008bayesian}. 

For the 2 dimensional Rosenbrock and Himmelblau functions, we obtain an average minimum ESS of $\sim100$ for $N_{\rm c} = 1000$. This suggests that, on average, it takes around 10 iterations to generate a new, independent sample. Empirically, we find this scales as $\sim 1/n_{\rm dim}$  for higher dimensions. 


\subsection{Fast-slow decorrelation}

In practice, the likelihood function can be computationally more expensive to evaluate for some parameters  (`slow' parameters) than others (`fast' parameters). In astronomy/cosmology applications, for example, nuisance parameters  are often much faster to evaluate than physical parameters of the model, when keeping physical parameters fixed. It is therefore desirable to split parameter space into a  speed hierarchy, allowing for fast exploration of subspaces where it is inexpensive to evaluate the likelihood~\cite{2013PhRvD..87j3529L}. 

\begin{figure}
\centering
\includegraphics[width=60mm, angle=0]{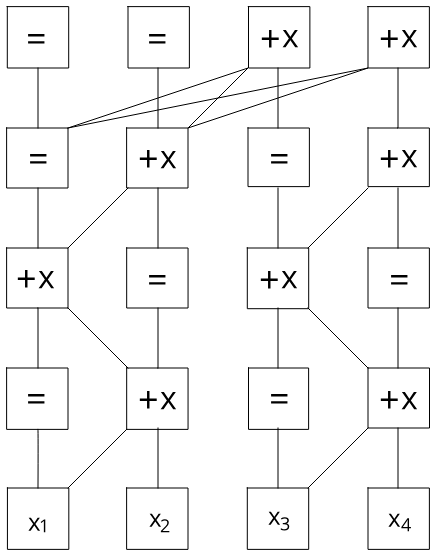}
\caption{\label{fig:fastslow} Illustration of transformations for a single fast-slow hierarchy, where $x_1$ and $x_2$ are slow parameters and $x_3$ and $x_4$ are fast parameters. The $+$ and $\times$ represent translation and scale operations  of the NNs respectively, and $=$ indicates the input is unmodified. A sequence of 3 transformations is applied to the slow and fast subspaces, then they are decorrelated by applying a further transformation. Changes to only fast parameters do not change slow parameters. 
 }
\end{figure}

Fast-slow decorrelation can naturally incorporated into our method by fitting flows to each subspace and performing a further transformation to decorrelate them. In the case of a single hierarchy, for example, we fit separate flows to the slow (${\bf x}_s$) and fast (${\bf x}_f$)  subspaces, concatenate the output into the vector (${\bf y}_s,  {\bf y}_f$), and then apply a transformation with mask $( {\bf 1}, {\bf 0})$. This means that slow parameters are unchanged by updating only the fast block, and a slow update changes both fast and slow parameters. This is illustrated in Fig.~\ref{fig:fastslow}.

Given a speed hierarchy, we choose the sampling rate to be proportional to the number of parameters in each block. In the case of a single hierarchy, with $n_{\rm dim} = n_{\rm slow} + n_{\rm fast}$,  where $n_{\rm slow}$ is the number of slow parameters and $n_{\rm fast}$ the number of fast parameters,  at each MCMC iteration we perform a fast update  with probability $n_{\rm fast} / ( n_{\rm slow} + n_{\rm fast} )$, otherwise performing a slow update. In our experiments we find this leads to a minimum ESS similar to the full update of all parameters.

\section{Neural Nest Algorithm} \label{sec:method}

In this section we give further details of our algorithm applied to nested sampling, although it could easily be incorporated into other MCMC frameworks. As outlined in section~\ref{sec:nested}, the algorithm begins by drawing $n_{\rm live}$  points from the prior distribution $\pi({\bf x})$. At each iteration the point  with the lowest likelihood (denoted by $\lambda_{\star}$) is replaced by a new sample drawn from the prior, subject to the condition that  $\mathcal{L} ({\bf x}) > \lambda_{\star}$. 

To obtain a new, independent sample, we first train a flow on the current set of live points.  Starting from an existing live point (chosen at random),  we then perform sampling in latent space, finally accepting the new point after $n_{\rm MCMC}$ steps, with the requirement that it must have made at least one move (in practice it will make many moves). Further pertinent details of our implementation are:

\begin{description}
\item[Number of MCMC iterations] Based on estimates of the ESS, we set $n_{\rm MCMC} = 5 n_{\rm dim}$.  Empirically, we find this works well for a range of target distributions. We monitor the ESS to ensure it does not significantly drop below 1 as the algorithm progresses, and that the chain performs a large number of updates by adjusting the proposal width as in~(\ref{eqn:proposalwidth}).
\item[Fast-slow hierarchy] We will consider models with either no hierarchy ($n_{\rm dim} = n_{\rm slow})$ or a single fast-slow hierarchy ($n_{\rm dim} = n_{\rm slow} + n_{\rm fast})$. In the latter case, we perform fast updates at each MCMC iteration with probability $n_{\rm fast} / ( n_{\rm slow} + n_{\rm fast} )$, otherwise performing a slow update that changes all parameters. On average, there will be approximately $5 n_{\rm slow}$ slow likelihood evaluations per chain. 
\item[Initial rejection sampling] In the initial stages of selecting a new point, the prior volume $X \sim 1$. In this case, it is more efficient to use rejection sampling from the prior hypercube. We switch to MCMC when the rejection efficiency is equal to the MCMC efficiency, i.e. after the prior volume has decreased by a factor of $1/(5 n_{\rm slow})$.
\item[Training updates] The set of live points changes relatively slowly, so we only retrain the flow every $n_{\rm live}$ iterations. We train each update for 50 epochs. Training is fast, taking $<60$ seconds on a CPU. The NN is trained by backpropagating
the loss in~(\ref{eqn:loss}) using the Adam optimiser~\cite{2014arXiv1412.6980K}.
\item[Adding jitter] We add `jitter' (random perturbations) to the set of live points during training to  reduce overfitting.  Jitter is chosen to be Gaussian with zero mean and a standard deviation of $0.2$ times the average nearest neighbour separation between live points. The level of jitter therefore reduces as the algorithm progresses.
\item[Validation data] During training we use $90\%$ of the current live points to train the flow. The remaining $10\%$ are used as validation data to ensure the loss does not increase due to overfitting. 
\item[Termination] The algorithm is terminated on determining the fractional remaining $Z$ to 0.5 in log-evidence (see~\cite{2008MNRAS.384..449F} for details). 
\item[Parallelisation] We parallelise our code using MPI, communicating the set of live points between processes. Each process trains a separate flow to the (same) set of  points, providing a type of ensembling across NNs.  A new live point is generated by each process, and these are communicated back to the master process, which is responsible for updating the set of live points. 
\end{description}

The NN was coded using the {\tt PyTorch} library\footnote{\href{https://pytorch.org/}{https://pytorch.org/}}  and the nested sampling code is available on request from the author. 


\section{ Results} \label{sec:results}

\subsection{Analytic likelihoods}

We first test our method on several challenging analytic likelihoods:

\begin{description}
\item[Mixture of 4 Gaussians] This is the same multi-modal distribution given in~\cite{2017arXiv170403459H}, with a likelihood function
\begin{equation}\label{equ:gaussian_mix}
    \mathcal{L}({\bf x}) = \sum_{m=1}^M W^{(m)} {\left(2 \pi {\bm \sigma^{(m)}}^2\right)}^{-d/2} \exp\left( -\frac{{| {\bf x} - {\bm \mu}^{(m)}|}^2}{2 {\bm \sigma^{(m)}}^2}\right).
\end{equation}
We also consider $M=4$, with weights $W^{(1)}=0.4$, $W^{(2)}=0.3$, $W^{(3)}=0.2$, $W^{(4)}=0.1$. The only non-zero components of ${\bm \mu}$ are $\mu^{(1)}_{2} = - \mu^{(2)}_{2} = \mu^{(3)}_{1}  = -\mu^{(4)}_{1}  = 4$. The standard deviation is ${\bm \sigma}^{(m)} = 1$ for all $m$. We choose a uniform prior of $\mathcal{U}(-10, 10)$ on the parameters ${\bf x}$. The analytic expression for the evidence is $\log Z = - n_{\rm dim}  \log 20$.
\item[Rosenbrock function] The is the archetypal example of a banana shaped degeneracy, with a log-likelihood
\begin{equation}
\log \mathcal{L}({\bf x}) = - \sum_{i=1}^{n_{\rm dim} - 1} \left[ \left(1 -x_i \right)^2 + 100 \left(x_{i+1}  - x_i^2 \right)^2 \right]\,.
\end{equation}
We choose uniform priors  $\mathcal{U}(-5, 5)$ on the parameters ${\bf x}$.  The analytic evidence for $n_{\rm dim}=2$ is $\log Z = - 5.80$~\cite{2012MNRAS.421..169G}. There is no analytic expression for $n_{\rm dim} > 2$, so  for  $n_{\rm dim}=3$ we perform  numerical integration to obtain the ground truth value. For higher dimensions we found this too expensive to compute numerically.
\item[Himmelblau function] This is an example of a multi-modal distribution, with a log-likelihood
\begin{equation}
\log \mathcal{L}({\bf x}) = -  \left( x_1^2 + x_2 - 11\right)^2 - \left(x_1 + x_2^2 - 7 \right)^2\,.
\end{equation}
We also choose uniform priors   $\mathcal{U}(-5, 5)$ on the parameters ${\bf x}$. The Himmelblau function has four identical local minima at (3, 2), (-2.81, 3.13), (-3.78, -3.28) and (3.58, -1.85). There is also no analytic expression for the evidence,  so we perform numerical integration to obtain the ground truth value for $Z$.
\end{description}

\begin{figure}
\centering
\includegraphics[width=88mm, angle=0]{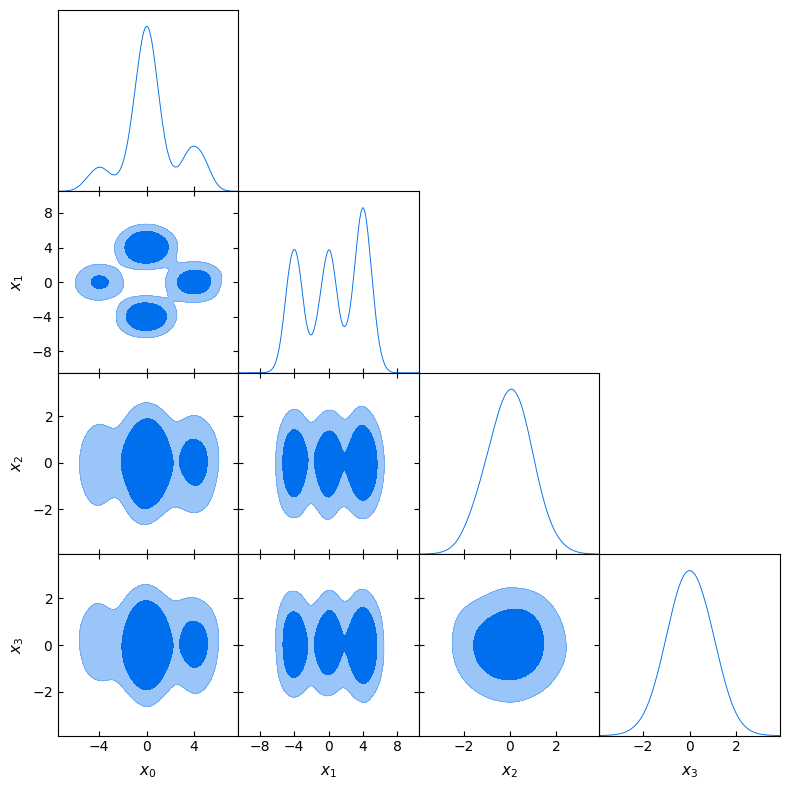}
\caption{\label{fig:mog4_corner} Marginalised 1 and 2-d posterior distributions  for the Gaussian mixture model with $n_{\rm dim}=4$.}
\end{figure}

In Fig.~\ref{fig:mog4_corner} we show the marginalised 1 and 2-d posterior distributions for the  Gaussian mixture model with $n_{\rm dim}=4$. These marginalised values  agree well with the expected values.  

In Tab.~\ref{tab:ncall} we show  $\log Z$ and the number of likelihood evaluations for each of the 3 analytic likelihoods. We compare our results  to the nested sampling codes {\tt MultiNest}~\cite{2009MNRAS.398.1601F} and {\tt PolyChord}~\cite{2015MNRAS.453.4384H}. {\tt MultiNest} uses multi-modal ellipsoidal rejection sampling,  and {\tt PolyChord} uses MCMC slice-sampling with multi-modal clustering. Each is run with their default settings\footnote{An efficiency factor of 0.3 is used for {\tt MultiNest} and $n_{\rm repeats}=5 n_{\rm dim}$  for  {\tt PolyChord}.}, and are set to stop on determining the fractional remaining log-evidence to an accuracy of 0.5.

 In our code, we use 5 transformations in the NVP flow, each parameterised by a fully connected  $s$ and $t$ network with an input layer of dimension 2, two hidden layers of dimension 128, and an output layer of dimension 2.  We use rectified non-linear activation (ReLU) functions after the input and hidden layers in each network. In all three codes we set $n_{\rm live} = 1000$
 and perform 5 separate runs to obtain summary statistics of $\log Z$ and the number of likelihood evaluations. Each code also produces an estimate of the evidence error for each run.

\begin{table*}[t!]
\begin{tabularx}{\textwidth}{ | Y | Y | Y | Y | Y | Y | Y |} \hline \hline
Likelihood & $n_{\rm slow}$ & $n_{\rm fast}$ & Ground truth &MultiNest~\cite{2009MNRAS.398.1601F} & PolyChord~\cite{2015MNRAS.453.4384H}  & Ours \\ \hline
  Gaussian mix. & 5 & 0  & -14.98 &  $-14.94 \pm 0.04$  & $-14.91 \pm 0.06$ &  $-14.91 \pm 0.05$ \\ 
 & & & & (34,459) & (949,706) & (139,755) \\ \hline
   Gaussian mix. & 10 &0  & -29.96 & $-29.90 \pm 0.06$  & $-29.97 \pm 0.10$ &  $-29.96 \pm 0.09$ \\ 
 & & & & (73,041) & (3,811,072) & (582,780) \\ \hline
   Gaussian mix. & 20 & 0  & -59.91 & $-59.21 \pm 0.12$  & $-59.91 \pm 0.22$ &  $-59.95 \pm 0.16$ \\ 
 & & & & (249,826) & (15,146,455) & (2,577,875) \\ \hline
    Gaussian mix. & 30 & 0  & -89.87 & $-88.42 \pm 0.09$  & $-89.85 \pm 0.33$  & $-89.79 \pm 0.10$  \\ 
 & & & &(697,089)  & (33,650,878)  &  (6,340,351) \\ \hline \hline
 
   Gaussian mix. & 2 & 3  & -14.98 &  N/A  &   &  $-15.01 \pm 0.04$ \\ 
 & & & &  &  & (58,460 ) \\ \hline
   Gaussian mix. & 2 & 8  & -29.96 & N/A &  & $-29.93 \pm 0.08$ \\ 
 & & & &  & & (121,668) \\ \hline
   Gaussian mix. & 2 & 18  & -59.91 &  N/A &  & $-59.80 \pm 0.17$ \\ 
 & & & &  &  & (261,827) \\ \hline
    Gaussian mix. & 2 & 28  & -89.87 &  N/A &  &  $-89.72 \pm 0.06$ \\ 
 & & & &   &   &  (425,564) \\ \hline \hline
 
  Himmelblau & 2 & 0 & $-5.54 \pm  0.02^{\star}$ & $-5.51 \pm 0.06$ & $-5.44 \pm 0.04$ &  $-5.48 \pm 0.08$ \\ 
 & & & & (21,110) & (259,506) & (47,880) \\ \hline \hline
 
 Rosenbrock & 2 & 0 & $-5.80$ & $-5.83 \pm 0.03$& $-5.82 \pm 0.12$  &  $-5.77 \pm 0.08$ \\ 
 & & & &(21,612) & (340,022 ) & (42,173) \\ \hline
Rosenbrock &  3 & 0 & $-10.46 \pm  0.03^{\star}$ & $-10.46 \pm 0.09$ &$-10.41 \pm 0.11$ . & $-10.44 \pm 0.13$  \\ 
& &  & &(37,658) &(775,309)  & (97,648) \\ \hline
Rosenbrock & 4 & 0 &  N/A & $-14.94 \pm 0.11$  & $-15.13 \pm 0.11$  & $-15.14 \pm 0.09$ \\ 
& & & &(58,606) & (1,443,530) & (195,704) \\ \hline
Rosenbrock & 5 & 0 & N/A& $-19.63 \pm 0.08$ & $-19.82 \pm 0.08$  & $-19.67 \pm 0.04$ \\ 
& & & & (78,346)& (2,308,802) & (319,325) \\ \hline
Rosenbrock & 10 & 0 & N/A& $-42.13 \pm 0.10$ &$-42.81 \pm 0.45$  & $-43.04 \pm 0.18$ \\ 
& & & & (385,446)& ( 9,742,368) & (1,468,468) \\ \hline
Rosenbrock & 20 & 0 & N/A & $-87.67 \pm 0.31$ & $-91.63 \pm 0.95$ & $-91.83 \pm 0.29$  \\ 
& & & & (6,459,763) & (38,047,353 ) & (6,803,067) \\ \hline
Rosenbrock & 30 & 0 & N/A& $-134.05$ &  $-138.79 \pm 1.74$  & $-141.81 \pm 0.28$ \\ 
& & & & (67,612,863) & (87,761,897 ) &  (16,305,276 ) \\ \hline 
 
\end{tabularx}
\caption{\label{tab:ncall}  Average $\log Z$ and number of slow likelihood evaluations for the analytic likelihoods. Values and errors are averaged over 5 runs.  Ground  truth values denoted by a $^{\star}$ were obtained by numerical integration. {\tt PolyChord} is also capable of a fast-slow hierarchy but we have not performed comparisons in this case.}
\end{table*}

For the Gaussian mixture model we obtain results consistent with the ground truth values. The number of required likelihood evaluations is around a factor of 5-10 higher (i.e. less efficient) than {\tt MultiNest}, primarily due to the $n_{\rm MCMC} = 5 n_{\rm dim}$ iterations we perform to obtain a new, independent sample. In contrast, the number of  evaluations required per sample for {\tt MultiNest} is only $\sim 3$. Using default settings,   however, {\tt MultiNest} tends to overestimate $\log Z$ for $n_{\rm dim} \geq 10$ -- this can be alleviated by decreasing the efficiency parameter, but at the cost of more likelihood evaluations.  Decreasing  the efficiency to 0.05, for example, gives $\log Z =  -59.92 \pm 0.07 $ for $n_{\rm dim}=20$, with  an average 703,182 likelihood evaluations, and  $\log Z =  -89.95 \pm 0.15 $ for $n_{\rm dim}=30$, with 962,111 likelihood evaluations. This means that {\tt MultiNest}  is still a factor $\sim 5$ times more efficient. Our error estimate in $\log Z$ using~(\ref{eqn:zerr}) is consistent with our summary statistics over 5 runs, being 0.08, 0.12, 0.17 and 0.21 for $n_{\rm dim}=5, 10, 20$ and 30 respectively. 

We also consider the same Gaussian mixture model but with a fast/slow  hierarchy. In this case we choose $x_1$ and $x_2$ to be slow parameters, with the remainder fast parameters. We take the number of likelihood evaluations to be the number of {\em slow evaluations}. This number  is now significantly reduced and is comparable to {\tt MultiNest} at low dimensions,  which does not implement any speed hierarchy,  and is more efficient at high dimensions.

For the Himmelblau function we obtain results consistent with the ground truth value. The number of likelihood evaluations is now only around a factor of 2 higher than {\tt MultiNest}. This is an improvement over the mixture model at the same dimension, as ellipsoidal rejection sampling is less efficient for non-Gaussian distributions.

For the Rosenbrock function we also obtain results consistent with the available ground truth values. For $n_{\rm dim}=2$  the number of likelihood evaluations is now less than a factor of 2 higher than {\tt MultiNest}, and for  $n_{\rm dim} \geq 20$ the poor scaling of rejection sampling becomes apparent. There is a difference in $\log Z$ compared to  {\tt MultiNest} for $n_{\rm dim} \geq 20$, and although the results of {\tt PolyChord} are consistent with ours, the lack of a ground truth value makes it difficult to draw conclusions.  Our error estimate in $\log Z$ using~(\ref{eqn:zerr}) is again consistent with our summary statistics, being 0.07, 0.09, 0.11, 0.13, 0.19, 0.28 and 0.34 for $n_{\rm dim}=2, 3, 4, 5, 10, 20$ and 30 respectively. 

Compared to  {\tt PolyChord}, also an MCMC sampler, our method requires  a lower number of likelihood evaluations, by a factor of $\sim 5$. {\tt PolyChord} also performs $n_{\rm MCMC} = 5 n_{\rm dim}$ repeats to obtain a new sample, but requires additional evaluations to determine the width of the slice. Recently, {\tt dyPolyChord}~\cite{2017arXiv170403459H} has been developed, which dynamically allocates live points during  sampling. We have not performed direct comparisons with  {\tt dyPolyChord} in Tab.~\ref{tab:ncall}, as we wish to compare the efficiency of each algorithm with a constant  number of live points. From our experiments, however, {\tt dyPolyChord} has similar performance to {\tt PolyChord} at low dimensions, but significantly improves the scaling at higher dimensions. For the Rosenbrock function, for example,  an average of 6,660,409 likelihood evaluations are required for  $n_{\rm dim}=20$ and only 9,182,957 for $n_{\rm dim}=30$.
 
Other MCMC based results in the literature are limited, but in~\cite{2013AIPC.1553..106F} it was shown that Galilean dynamics required around 120,000 and 220,000 likelihood evaluations for the Himmelblau and $n_{\rm dim}=2$ Rosenbrock functions respectively.


\subsection{Planck}

Our {\tt Python} implementation can easily be integrated with codes such as {\tt MontePython}~\cite{Audren:2012wb, Brinckmann:2018cvx} and {\tt cobaya}\footnote{\href{https://github.com/CobayaSampler/cobaya}{https://github.com/CobayaSampler/cobaya}} to perform cosmological  parameter estimation and model selection. The  {\em Planck} datasets used in our analysis come from the 2015 mission~\cite{Aghanim:2015xee, Ade:2015xua}. In particular, we use the TT+lowP+lensing combination, which contains the 100-GHz, 143-GHz, and 217-GHz binned half-mission TT cross-spectra for $\ell=30-2508$ with CMB-cleaned 353-GHz map, CO emission maps, and {\em Planck} catalogues for the masks and 545-GHz maps for the dust residual contamination template. It also uses the joint temperature and the E and B cross-spectra for $\ell=2-29$ with E and B maps from the 70-GHz LFI full mission data and foreground contamination determined by 30-GHz Low Frequency Instrument (LFI) and 353-GHz High Frequency Instrument maps. The {\em Planck} lensing likelihood~\cite{Ade:2015zua} uses both temperature and polarization data in the multipole range $\ell=100-2048$ to estimate the lensing power spectrum.

We use the full version of the {\em Planck} likelihood with  {\tt MontePython},  fitting for a total of 6 base LCDM parameters and 15 nuisance parameters.  We also fit for simple one-parameter extensions to LCDM with a variable effective number of neutrino species $N_{\rm eff}$ and curvature density $\Omega_{\rm K}$.  We assume uniform priors on the cosmological parameters, with the upper and lower limits  corresponding to  approximate {\em Planck} $\pm 5 \sigma$ values.  To account for any Gaussian priors on nuisance parameters used in the {\em Planck} analysis,  we use uniform priors with  $\pm 5 \sigma$ limits, and add an additional term to the likelihood function. The prior ranges, along with a description of each parameter, are shown in Tab.~\ref{tab:planck_priors}. 

\begin{table*}
\begin{center}
\centering
\begin{tabularx}{\textwidth}{ | s | s | s | b |  } \hline \hline
Lower &Parameter  & Upper & Description \\ \hline
$0.0211$ & $\leq \Omega_{\rm b} h^2 \leq$ &$0.0234$ & Physical baryon density  \\
$0.109$ & $\leq \Omega_{\rm c} h^2 \leq$ &$0.131$ & Physical CDM density   \\
$1.038$ & $\leq 100 \Theta_{\rm s} \leq$ &$1.044$ &  Ratio of angular diameter distance to sound horizon   \\ 
$2.91$  & $\leq \ln \left( 10^{10} A_{\rm s} \right) \leq$ &$3.27$ & Scalar amplitude   \\ 
$0.93$  & $\leq n_{\rm s} \leq$ &$1.0$ & Scalar spectral index  \\
$0.05$ & $\leq \tau \leq$ &$0.15$ & Optical depth to reionization  \\ \hline 
$1.5$ & $\leq N_{\rm eff} \leq$ &$4.5$ & Effective number of neutrinos    \\ 
$-0.1$ & $\leq \Omega_{\rm K} \leq$ &$0.05$ & Curvature density    \\ \hline 
$0\,{\mu}{\rm K}^2$ & $\leq A^{\mathrm{CIB}}_{217}   \leq$ &$200\,{\mu}{\rm K}^2$ & CIB amplitude at 217 GHz   \\ 
$0\,{\mu}{\rm K}^2$  & $\leq A^{\mathrm{kSZ}}  \leq$ &$10\,{\mu}{\rm K}^2$  & kSZ amplitude at 143 GHz  \\ 
$0\,{\mu}{\rm K}^2$  & $\leq A^{\mathrm{tSZ}}_{143}   \leq$ &$10\,{\mu}{\rm K}^2$  & tSZ amplitude at 143 GHz  \\ 
$0$ & $\leq \xi^{\mathrm{tSZ}\times\mathrm{CIB}}  \leq$ &$1$ & tSZ-CIB template amplitude  \\ 
$0\,{\mu}{\rm K}^2$ & $\leq A^{\mathrm{PS}}_{100}  \leq$ &$400\,{\mu}{\rm K}^2$ & Point source amplitude at 100 GHz   \\ 
$0\,{\mu}{\rm K}^2$ & $\leq A^{\mathrm{PS}}_{143}   \leq$ &$400\,{\mu}{\rm K}^2$ & Point source amplitude at 143 GHz  \\ 
$0\,{\mu}{\rm K}^2$ & $\leq A^{\mathrm{PS}}_{143\times 217}  \leq$ &$400\,{\mu}{\rm K}^2$ & Point source amplitude at 143x217 GHz   \\ 
$0\,{\mu}{\rm K}^2$ & $\leq A^{\mathrm{PS}}_{217}  \leq$ &$400\,{\mu}{\rm K}^2$ & Point source amplitude at 217 GHz  \\ 
$0\,{\mu}{\rm K}^2$ & $\leq A^{{\rm dust}TT}_{100} \leq$ &$17\,{\mu}{\rm K}^2$ & Dust amplitude at 100 GHz   \\ 
$0\,{\mu}{\rm K}^2$ & $\leq A^{{\rm dust}TT}_{143} \leq$ &$19\,{\mu}{\rm K}^2$ & Dust amplitude at 143 GHz  \\ 
$0\,{\mu}{\rm K}^2$ & $\leq  A^{{\rm dust}TT}_{143\times 217} \leq$ &$63.5\,{\mu}{\rm K}^2$ & Dust amplitude at 143x217 GHz   \\ 
$0\,{\mu}{\rm K}^2$ & $\leq A^{{\rm dust}TT}_{217}  \leq$ &$180\,{\mu}{\rm K}^2$ & Dust amplitude at 217 GHz  \\ 
$0.994$ & $\leq c_{100}  \leq$ &$1.004$ & Calibration factor for 100/143 GHz  \\ 
$0.985$ & $\leq c_{217}  \leq$ &$1.005$ & Calibration factor for 217/143 GHz  \\ 
$0.9875$ & $\leq y_{\rm cal} \leq$ &$1.0125$ & Total {\em Planck} calibration  \\ \hline
\end{tabularx}
\caption{Prior ranges for the base LCDM (top), one-parameter extensions (middle) and nuisance parameters (bottom), together with the resulting posterior values.} 
\label{tab:planck_priors}
\end{center}
\end{table*}

\begin{table*}
\begin{center}
\centering
\begin{tabularx}{\textwidth}{| Y | Y | Y | Y  | } \hline \hline
Parameter  & Base LCDM  &  $+N_{\rm eff}$ & $+\Omega_{\rm K}$  \\ \hline
 $ \Omega_{\rm b} h^2 $ & 			$0.02229_{-0.00024}^{+0.00023}$ & 	$0.02237_{-0.00033}^{+0.00033}$ &	 	$0.02233_{-0.00027}^{+0.00024}$ \\
$ \Omega_{\rm c} h^2 $  &  			$0.1182_{-0.0020}^{+0.0019}$  & 		$0.1190_{-0.0036}^{+0.0042}$ &  		$0.1176_{-0.0024}^{+0.0022}$	\\
 $ 100 \Theta_{\rm s} $ & 				$1.0420_{-0.0004}^{+0.0004}$ & 		$1.0420_{-0.0007}^{+0.0007}$ &  		$1.0420_{-0.0005}^{+0.0005}$	\\ 
 $ \ln \left( 10^{10} A_{\rm s} \right) $  & 	$3.078_{-0.032}^{+0.020}$ & 			$3.084_{-0.041}^{+0.030}$ & 			$3.070_{-0.036}^{+0.021}$	\\ 
 $ n_{\rm s} $ & 					$0.9690_{-0.0061}^{+0.0055}$ & 		$0.9726_{-0.013}^{+0.014}$ & 			$0.9708_{-0.0069}^{+0.0066}$	\\
 $ \tau $  & 						$0.073_{-0.018}^{+0.010}$  & 			$0.076_{-0.020}^{+0.012}$ & 			$0.0703_{-0.020}^{+0.006}$	\\ \hline 
 $ N_{\rm eff} $   & 					$3.046$  & 						$3.11_{-0.29}^{+0.30}$  & 				$3.046$  \\ 
 $ \Omega_{\rm K} $ & 				$0.0$ & 							$0.0$ &  							$-0.0042_{-0.0073}^{+0.0089}$ \\ \hline 
\end{tabularx}
\caption{Marginalised values for the  base LCDM,  ${\rm LCDM}+N_{\rm eff}$  and ${\rm LCDM}+\Omega_{\rm K}$ cosmological parameters.} 
\label{tab:planck_parameters}
\end{center}
\end{table*}

\begin{figure*}
\centering
\includegraphics[width=\textwidth, angle=0]{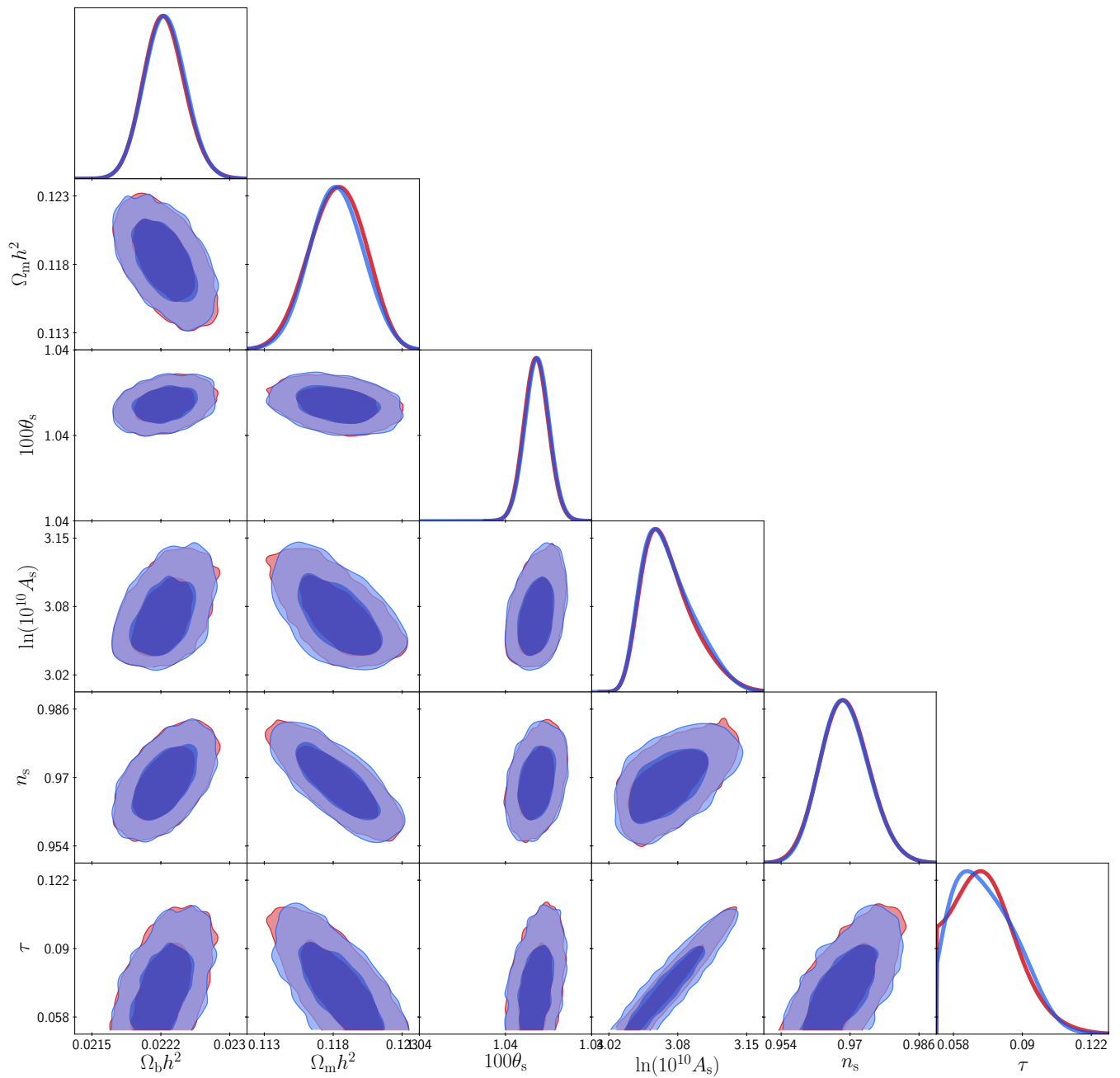}
\caption{\label{fig:planck_corner} Marginalised 1 and 2-d posterior distributions  for the base LCDM cosmological parameters. In red we show results from our nested sampling method, and in blue results from standard parameter estimation using Metropolis-Hastings MCMC.}
\end{figure*}

For {\em Planck} the total computational time is dominated by the calculation of the cosmological observables, so we use a fast hierarchy for the nuisance parameters. We use  $n_{\rm live} = 500$ points and the same network architecture as in the previous section. In Tab.~\ref{tab:planck_parameters} we give the marginalised  cosmological parameters for the base LCDM,  ${\rm LCDM}+N_{\rm eff}$  and ${\rm LCDM}+\Omega_{\rm K}$ models. These agree very well with the published {\em Planck} values, and in Fig.~\ref{fig:planck_corner} we show marginalised 1 and 2-d posterior distributions for the base LCDM model, compared to results from standard MCMC.   These  again agree extremely well, showing that we obtain accurate parameter constraints using our nested sampler.

We have also calculated the evidence, finding the Bayes factor to be $\log B = - 1.7 \pm 0.2$ and $-2.1 \pm 0.2$ for  ${\rm LCDM}+N_{\rm eff}$ and  ${\rm LCDM}+\Omega_{\rm K}$ respectively.  The error on the Bayes factor is obtained from adding the errors from~(\ref{eqn:zerr}) in quadrature. In the revised Jeffreys scale~\cite{kass1995bayes}, $|\log B| > 1$ is regarded as positive evidence, $|\log B| > 3$ as strong evidence, and $|\log B| > 5$  as very strong. These results therefore suggest that {\em Planck} disfavours both  extensions to LCDM. Although the evidence is dependant on the choice of priors, our results are consistent with those in~\cite{2017PhRvL.119j1301H}, who reuse MCMC chains produced for parameter inference to calculate the evidence. 
        
\section{Conclusions} \label{sec:conclusions}

In this paper we have trained a neural network to parameterise efficient MCMC proposals, by transforming the target distribution to a simpler latent representation. This approach is inspired by representative learning, which suggests that deep NNs yield latent spaces in which Markov chains can mix faster. Our method is a non-linear extension of the commonly used technique of transforming  parameter space using the  Cholesky decomposition of the covariance matrix.

We have applied this method to nested sampling,  finding excellent performance on highly curved and multi-modal targets. At low dimensions the efficiency is within a factor of a few times that of state-of-the-art multi-modal rejection sampling, but has better scaling in higher dimensions. Parameter space can also naturally be split into a speed hierarchy, making it suitable for models with a subset of parameters where it is inexpensive to evaluate the likelihood. We demonstrate this for {\em Planck} data in $\sim20$ dimensional parameter space, accurately recovering the expected posterior distributions. As example applications, we calculate the Bayesian evidence for variable effective number of neutrino species $N_{\rm eff}$ and curvature density $\Omega_{\rm K}$, finding the data disfavours these extensions to LCDM.

There are several possibilities for future work. Firstly, it would be interesting to see if the flow model can more naturally be extended to multi-modal distributions. Currently, the latent representation forms narrow connecting ridges between modes, which reduces the efficiency on models with a very high number of modes. One could also potentially use more general types of neural network, but these may not have the desirables properties of being invertible with tractable  Jacobian determinants.

In terms of nested sampling, it has recently been shown that dynamically allocating the number of live points can significantly improve performance. It would be interesting to apply this technique to our method, potentially even using a neural network to estimate the posterior mass $\mathcal{L}(X) X$ and control the allocation of points.

In follow up work we will develop a NN sampler specifically designed for fast inference, that can easily be integrated into standard parameter estimation codes. This would improve on  the standard technique of using the covariance matrix to parameterise the proposal function, working for  both highly curved and multi-modal likelihoods. With the ability of NNs to characterise complex data by simple representations, we expect they will become useful tools to improve the speed of inference on a variety of problems.\\


\section*{Acknowledgements}

We appreciate helpful conversations with Steven Bamford, Simon Dye, Juan Garrahan and Dominic Rose, and Will Handley for very useful comments on the nested sampling algorithm. AM is supported by a Royal Society University Research Fellowship.

\bibliography{bib}

\end{document}